\UseRawInputEncoding

\documentclass[fleqn,usenatbib]{mnras}
\usepackage{newtxtext,newtxmath}
\usepackage[T1]{fontenc}
\usepackage{lineno}
\DeclareRobustCommand{\VAN}[3]{#2}
\let\VANthebibliography\thebibliography
\def\thebibliography{\DeclareRobustCommand{\VAN}[3]{##3}\VANthebibliography}

\usepackage{graphicx}	% Including figure files
\usepackage{amsmath}
\usepackage{amssymb}	% Extra maths symbols
\usepackage{multicol}        % Multi-column entries in tables
\usepackage{bm}		% Bold maths symbols, including upright Greek
\usepackage{pdflscape}	% Landscape pages
\usepackage{threeparttable}
\usepackage{multirow}
\usepackage{subfigure}
\newcommand{\kms}{km\,$\rm s^{-1}$}

\title[Probing the Galactic halo with RR Lyrae stars ]{Probing the Galactic halo with RR Lyrae stars $-$ IV. On the Oosterhoff dichotomy of RR Lyrae stars}

\author[Zhang et al.]{
Shan Zhang,$^{1,2,}$
Gaochao Liu,$^{1,2,}$\thanks{E-mail: gcliu@ctgu.edu.cn ; huangyang@ucas.ac.cn}
Yang Huang,$^{3,4,}$\footnotemark[1]
Zongfei Lv,$^{5,6,}$
Sarah Ann Bird,$^{1,2}$
Bingqiu Chen,$^{7}$
\newauthor
Huawei Zhang,$^{8,9}$ 
Timothy C. Beers,$^{10}$
Xinyi Li,$^{7}$
Haijun Tian,$^{11}$
 and Peng Zhang$^{1,2}$
\\
$^{1}$College of Science, China Three Gorges University, Yichang 443002, People's Republic of China\\
$^{2}$Center for Astronomy and Space Sciences, China Three Gorges University, Yichang 443002, People's Republic of China\\
$^{3}$School of Astronomy and Space Science, University of Chinese Academy of Sciences, Beijing 100049, People’s Republic of China\\
$^{4}$CAS Key Lab of Optical Astronomy, National Astronomical Observatories, University of Chinese Academy of Sciences, Beijing 100101, People’s Republic of China\\
$^{5}$Purple Mountain Observatory, University of Chinese Academy of Sciences, Nanjing, 210023, People's Republic of China\\
$^{6}$School of Astronomy and Space Sciences, University of Science and Technology of China, Hefei 230026, People's Republic of China\\
$^{7}$South-Western Institute For Astronomy Research, Yunnan University, Kunming 650500, People's Republic of China\\
$^{8}$Department of Astronomy, School of Physics, Peking University, Beijing 100871, People's Republic of China\\
$^{9}$Kavli Institute for Astronomy and Astrophysics, Peking University, Beijing 100871, People's Republic of China\\
$^{10}$Department of Physics and Astronomy and JINA Center for the Evolution of the Elements (JINA-CEE), University of Notre Dame, Notre Dame, IN 46556, USA\\
$^{11}$School of sciences, HangZhou Dianzi University, HangZhou 310018, People's Republic of China
}

\date{Accepted XXX. Received YYY; in original form ZZZ}

\pubyear{2023}

\begin{document}
\label{firstpage}
\pagerange{\pageref{firstpage}--\pageref{lastpage}}
\maketitle

\begin{abstract}
 We use 3653 (2661 RRab, 992 RRc) RR Lyrae stars (RRLs) with 7D (3D position, 3D velocity, and metallicity) information selected from SDSS, LAMOST, and Gaia EDR3, and divide the sample into two Oosterhoff groups (Oo\,I and Oo\,II) according to their amplitude-period behaviour in the Bailey Diagram.
 We present a comparative study of these two groups based on chemistry, kinematics, and dynamics.
 We find that Oo\,I RRLs are relatively more metal rich, with predominately radially dominated orbits and large eccentricities, while Oo\,II RRLs are relatively more metal poor, and have mildly radially dominated orbits. 
 %Other physical parameters of Oo\,I and Oo\,II, such as space and velocity distribution, show no difference. 
 The Oosterhoff dichotomy of the Milky Way's halo is more apparent for the 
 inner-halo region than for the outer-halo region.
 Additionally, we also search for this phenomenon in the halos of the two largest satellite galaxies, the Large and Small Magellanic clouds (LMC, SMC), and compare over different bins in metallicity. 
 We find that the Oosterhoff dichotomy is not immutable, and varies based on position in the Galaxy and from galaxy-to-galaxy. 
 We conclude that the Oosterhoff dichotomy is the result of a combination of stellar and galactic evolution, and that it is much more complex than the dichotomy originally identified in Galactic globular clusters. 
\end{abstract}


%%%%%%%%%%%%%%%%%%%%%%%%%%%%%%%%%%%%%%%%%%%%%%%%%%

%%%%%%%%%%%%%%%%% BODY OF PAPER %%%%%%%%%%%%%%%%%%

\section{Introduction}
\label{sec:intro}
RR Lyrae stars (RRLs) are large-amplitude radially pulsating variable stars, populating the intersection of the instability strip with the Horizontal Branch (HB) \citep{kolenberg2010depth,catelan2015pulsating,wang2021asteroseismology}. 
The majority of RRLs are classified based on their periods of oscillation and the amplitude and skewness of their light curves. The RRab stars have longer periods ($\sim0.45$ to $\sim1$ days) and larger amplitude, are pulsating in the radial fundamental mode, and exhibit asymmetric light curves. In contrast, RRc stars have shorter periods ($\sim0.25$ to $\sim0.45$ days) and generally lower amplitudes; they oscillate in the first-overtone mode and exhibit sinusoidal variations \citep{aerts2010asteroseismology,catelan2015pulsating,monelli2022rr}.

The Oosterhoff dichotomy is one of the most discussed long-term astrophysics problems, ever since \citet{oosterhoff1939some} pointed out that the RRLs in Galactic globular clusters can be split into two different groups: Oosterhoff I (Oo\,I) and Oosterhoff II (Oo\,II).
The Oo\,I RRLs have mean periods of $\langle P_{ab}\rangle\simeq0.56$ days for RRab stars and $\langle P_{c}\rangle\simeq0.31$ days for RRc stars. The mean periods of Oo\,II RRLs is longer than for the Oo\,I class; $\langle P_{ab}\rangle\simeq0.66$ days and $\langle P_{c}\rangle\simeq0.36$ days. \citet{arp1955colour} and \citet{preston1959spectroscopic} carried out spectroscopic studies for both classes, and found that Oo\,I RRLs are associated with globular clusters that are more metal rich and slightly younger than for Oo\,II RRLs.

Numerous investigations have been carried out over the past five decades to quantify the differences between the Oosterhoff groups, and attempt to explain the origin of its dichotomy. We summarise the leading explanations of the dichotomy here.

One possible origin is linked to stellar evolution and the nature of the pulsations.
Early studies showed that physical parameters, such as effective temperature, luminosity, mass, and the mode of the pulsation, can significantly influence the periods of RRLs.
\citet{1973On} conjectured that the temperature difference induced by the "hysteresis effect" is the reason for the Oosterhoff dichotomy.
\citet{sandage1981evidence,sandage1981oosterhoff} discovered a period shift at fixed temperature, which led him to conclude that the dichotomy is caused by differences in luminosity.
Using synthetic models of the HB, \citet{lee1990horizontal} showed that the evolution
away from the zero-age HB can explain the luminosity difference, which plays a role in the Oo\,II-group RRLs by increasing the mean luminosity and amplifying the dichotomy.

The second interpretation is related to the formation history of the two Oosterhoff groups.
According to the relative age estimation and the kinematics of RR Lyrae stars \citep{1993Globular,1999RR}, it has been suggested that the Oo\,II group was formed very early in the proto-Galaxy, while the Oo\,I group was formed at a later time and may be connected with merger events. 
\citet{jang2015star} proposed that the population-shift effect within the instability strip induced the period dichotomy for the globular clusters in the inner-halo and outer-halo regions, and that the
origin of the Oosterhoff dichotomy could arise from two or three distinct star-formation episodes. 

A third possibility is that the dichotomy is primarily an environmental effect. \citet{bono1994oosterhoff}, \citet{2009Astrophysics}, \citet{smith2011rr}, and \citet{2014Weak}
investigated nearby dwarf galaxies, and found that RRLs in Local Group galaxies and their globular clusters have intermediate Oosterhoff properties -- filling in the Oosterhoff gap with mean periods of 0.58 $\sim$ 0.62 days. This suggests that the Oosterhoff dichotomy found in the Milky Way (MW) does not exist in dwarf galaxies and their globular clusters. 
\citet{fabrizio2019use} collected a large data set of field RRLs with homogeneously determined metallicities, and concluded that the Oosterhoff dichotomy is mainly the consequence of the lack of intermediate-metallicity globular clusters in the MW.
It follows that the Oosterhoff dichotomy is not importantly connected with any RRL hysteresis effect \citep{castellani1983old,renzini1983current}.
\citet{fabrizio2021use} enlarged their data set and also analysed the first-overtone variables. The smooth distributions over the entire metallicity
range found for the pulsation periods and amplitudes of both RRab and RRc stars provided supporting evidence for this view.

RRLs, low-mass core helium-burning stars, undergo significant changes in effective temperature and surface gravity during their evolution \citep{catelan2015pulsating,Bono2020OnTM,monelli2022rr}. These changes result in variations in both pulsation period and luminosity amplitude, leading to discernible differences in periods between the two Oosterhoff groups.
Here, we propose that the Oosterhoff dichotomy is more complicated than the original definition based on MW globular clusters. 
Differences in the metal abundance of RRLs in the halo field of the MW primarily arise from the nature of the environment in which they form, prior to being deposited into the halo field through merger events and/or stripping due to tidal interactions.  

Recent studies \citep{Lancaster2018TheHA,Liu2022ProbingTG,wu2022contribution} have shown that the Gaia-Sausage-Enceladus (GSE; \citealt{Belokurov2018CoformationOT,Helmi2018TheMT}) substructure dominates the inner-halo region (IHR) of the MW, and has little effect on the outer-halo region (OHR) or satellite dwarf galaxies \citep{fabrizio2021use}. As a result, the Oosterhoff dichotomy is apparent in the IHR while it is not found in the OHR. We also explore this dichotomy for the LMC and SMC, the largest two satellite galaxies of the MW, and find that the presence of the Oosterhoff dichotomy varies across different galaxies. In addition, the dichotomy apparently differs for stars of different metallicity in the MW halo.
Moreover, there exist no significant differences in the basic properties of the Oo\,I and Oo\,II groups, such as their velocity distribution, spatial distribution, etc.
Hence, we favour the hypothesis that the Oosterhoff dichotomy is the result of a combination of stellar evolution and galaxy evolution.

\begin{figure*}
	\centering
	\includegraphics[width=1.6\columnwidth]{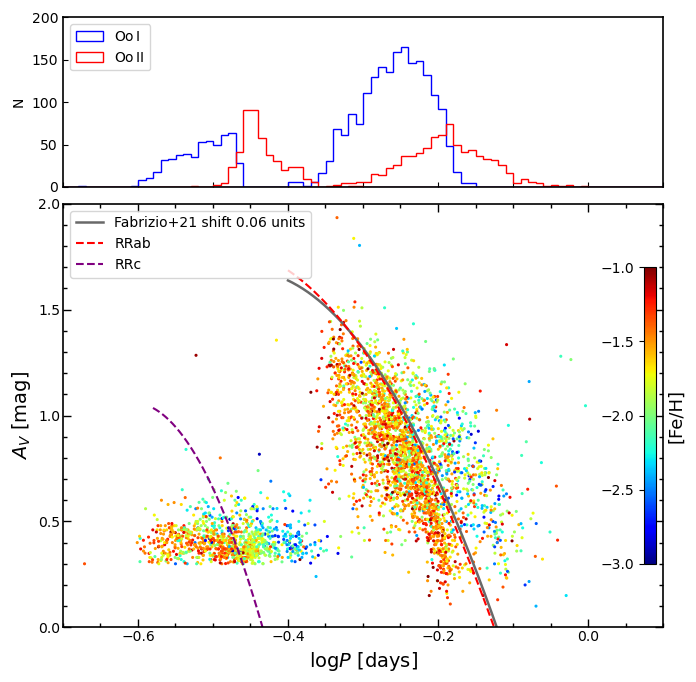}
	\caption{Period distribution of the Oosterhoff groups in the MW halo. The Bailey Diagram (visual amplitude vs. logarithmic period) is shown, colour-coded by metallicity as indicated by the colour bar. The grey solid line for RRab stars is from \citet{fabrizio2021use}. The red and purple dashed lines are used to divide the Oo\,I and Oo\,II groups for RRab and RRc stars from the polynomial regression method described in the text. The top histograms show the distribution of logP for the Oo\,I stars (blue) and Oo\,II stars (red). }
	\label{fig.Bailey}
\end{figure*}

In this paper, we present the kinematics, dynamics, and chemistry of the Oosterhoff groups in the Galactic halo using RRLs collected from large-scale
spectroscopic and photometric surveys. In Section\,\ref{sec:data}, we describe the RRL sample and separate it into the two Oosterhoff groups using the Bailey Diagram. 
The properties of the Oosterhoff groups are described in Section\,\ref{sec:result}. Section\,\ref{sec:dis}  presents a discussion and comparison with recent results published in the literature. We summarise 
in Section\,\ref{sec:sum}.

\section{Data}\label{sec:data}
\begin{figure}
	\centering    %??
	\subfigure
	{
		\begin{minipage}{7cm}
			\centering          
			\includegraphics[scale=0.46]{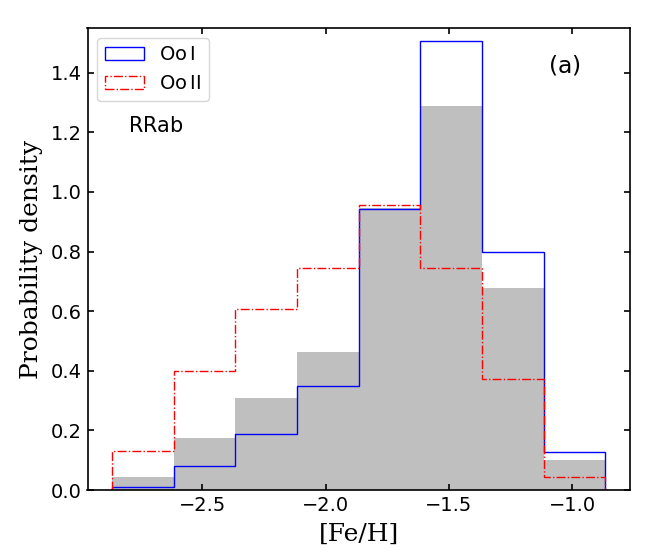}   
		\end{minipage}
	}
	\subfigure
	{
		\begin{minipage}{7cm}
			\centering      %????
			\includegraphics[scale=0.46]{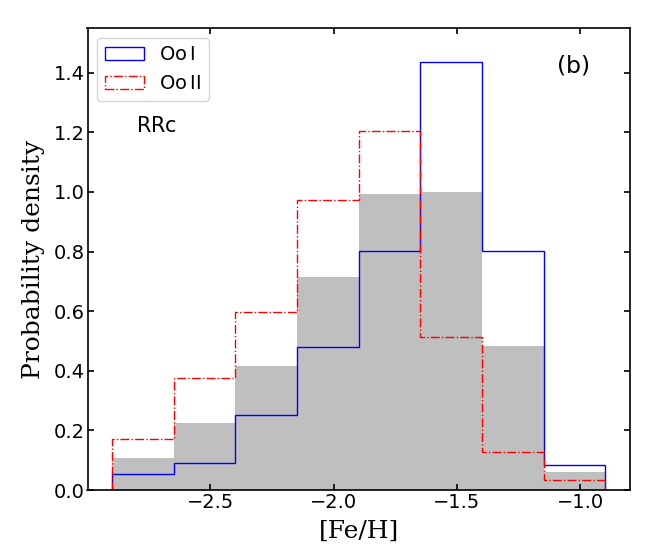}   
		\end{minipage}
	}
	\caption{Panel (a): The metallicity distribution of RRab variables is shown in the grey histogram ($\rm \langle [Fe/H] \rangle_{ab}=-1.67$). The blue and red histograms represent the metallicity distribution of Oo\,I and Oo\,II groups ($\rm \langle [Fe/H] \rangle_{Oo\,I}=-1.59$, $\rm \langle [Fe/H]\rangle_{Oo\,II}=-1.87$). Panel (b): The metallicity distribution of RRc variables ($\rm \langle [Fe/H] \rangle_{c}=-1.80$, $\rm \langle [Fe/H] \rangle_{Oo\,I}=-1.65$, $\rm \langle [Fe/H]\rangle_{Oo\,II}=-1.96$.)} 
	\label{fig.feh}  
\end{figure}

\subsection{The RR Lyrae Sample}
\label{subsec:RRl}

We assemble our sample of RRLs from two previously published catalogues.
The majority our our sample is drawn from the catalogue of \citet{Liu2020ProbingTG}, who collected 6268 unique RRLs by combining the Sloan Digital Sky Survey\,/\,Sloan Extension for Galactic Understanding and Exploration \citep[SDSS/SEGUE;][]{yanny2009segue} and the Large Sky Area Multi-Object Fiber Spectroscopic Telescope \citep[LAMOST;][]{deng2012lamost,zhao2012lamost,liu2014k} spectroscopic data with photometric data from the literature. They measured metallicities of 5290 RRLs using a least-$\chi^2$ fitting technique. By fitting empirical template radial velocity curves of RRLs, they estimated the systemic radial velocity for 3642 RRLs, and using the $PMZ$ (period, absolute magnitude, metallicity) or $M_{\rm V}$$-$ [Fe/H] relations, they obtained distance estimates for 4919 RRLs.

The second catalogue is from \citet{2022Probing}, who enlarged their data set by combining the recently published RRL catalogue from photometric surveys, such as {\it Gaia} \citep{2019Gaia}, the All-Sky Automated Survey for SuperNovae \citep[ASAS-SN;][]{jayasinghe2018asas}, Pan-STARRSI \citep{chambers2016pan,sesar2017machine}, and spectroscopic data from LAMOST and SDSS \citep{alam2015}. 
They employed the same methods as \citet{Liu2020ProbingTG} to estimate systemic radial velocities, metallicities, and distances for the RRLs. Both data sets use the shape of the light curve obtained from photometry to distinguish between RRab and RRc stars.

The total number RRLs we have for this study is 8172. By cross-matching with {\it Gaia} EDR3 \citep{GaiaCollaboration2021}, we obtain proper motions and {\it G}-band amplitudes for these stars. In order to study the Oosterhoff dichotomy of the Galactic halo, we select sources with [Fe/H] $\leq -1.0$, and $|Z| \ge 2$ \,kpc (which excludes potential contaminators from the Galactic disk and bulge).
We then exclude the stars without radial velocity, metallicity, proper motions, and amplitude measurements, and obtain a final sample of 2661 RRab stars and 992 RRc stars with full 7D information (3D position, 3D velocity, and metallicity).

\subsection{Coordinate System}
\label{subsec:coor}

We adopt two coordinate systems for this work.
One is a right-handed coordinate system $(X , Y , Z )$, where $X$ points in the direction opposite the Sun, $Y$ is in the direction of Galactic rotation, and $Z$ is towards the North Galactic Pole. The other is the Galactocentric spherical system $(r , \theta , \phi)$, where $r$ is the Galactocentric distance, $\theta$ increases from 0 to $\pi$ from the North Galactic Pole to the South Galactic Pole, and $\phi$ represents the azimuthal angle. Moreover, we adopt the distance from the Sun to the Galactic Centre as 8.34 kpc \citep{Reid2014TRIGONOMETRICPO}, the Solar motion as $(U_\odot,V_\odot,W_\odot)=(9.58, 10.52, 7.01)$\,\kms\ \citep{tian2015}, and the circular speed of the Local Standard of Rest as $240\,\rm km\,s^{-1}$ \citep{Reid2014TRIGONOMETRICPO}. We note that the choices of other values of the Solar motions and circular speed 
(e.g., \citealt{2015huang,2016huang,zhou2023}) at the Solar position have minor effects on our main results.

\subsection{Bailey Diagram}\label{subsec:bailey}

The Bailey Diagram (luminosity amplitude vs. period) is a useful diagnostic to investigate the properties of RRLs for two crucial reasons \citep{fabrizio2019use}. One is that the Bailey Diagram relies on two observables that are independent of distance and reddening. The other is that globular cluster RRLs can be divided into two groups, Oo\,I (short periods and metal rich) and Oo\,II (long periods and metal poor). The Oo\,II stars have slightly larger periods than Oo\,I stars when compared at similar amplitude. 
We note a possible exception that, in the case of the two metal-rich globular clusters NGC 6388 and NGC 6441, they are characterised by higher metallicity and longer mean periods. They are inconsistent with the typical pattern of decreasing periods with increasing metallicities of the Galactic globular clusters, and have thus been considered as the prototype of the third Oosterhoff (Oo\,III) groups \citep{Pritzl2000,2009Astrophysics,bhardwaj2020,bhardwaj2022near}. However, this is beyond the scope of this study, since we mainly focus on the metal-poor RRLs with [Fe/H]\,$\le -1.0$.

The Bailey Diagram enables one to split fundamental and first-overtone RRLs. 
The RRc variables are located in the low-amplitude, short-period (SP) region and their amplitude-period relationship exhibits a ``bell-shaped" or ``hairpin" distribution \citep{Petroni2003FieldAC,mcwilliam2011rr,fiorentino2017weak}. The RRab variables dominate the large-amplitude, long-period (LP) region, and exhibit a steady decrease in amplitude \citep{Bono2020OnTM}. The transition period between RRc to RRab stars covers a range $\log{P}$ $\sim$ $[-0.35, -0.30]$.
The RRL distribution in the Bailey Diagram mainly depends on their intrinsic parameters (stellar mass, luminosity, effective temperature, and chemical composition;  \citealt{bono2007rr,Bono2020OnTM}). 
The lower panel of Fig.\,\ref{fig.Bailey} shows the Bailey Diagram for our final sample colour-coded by metallicity (increasing [Fe/H] from blue to red). The RRab stars are located on the right side of this panel, and the RRc stars are on the left. We can clearly observe that RRab and RRc stars become more metal rich when moving from the LP to the SP regime, at fixed amplitude.

\begin{figure*}
	\centering
	\includegraphics[scale=0.7]{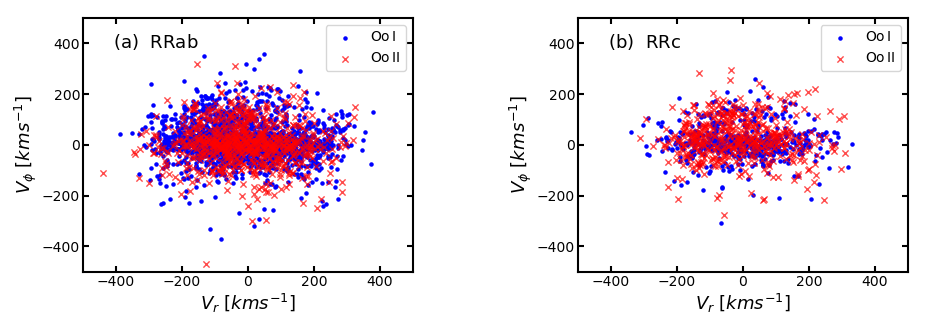}
	\caption{Panels (a) and (b) show the velocity components for RRab and RRc stars, respectively. The blue points and red crosses represent the Oo\,I and Oo\,II types, respectively.}
	\label{fig.vrvphi}
\end{figure*}

To delineate the salient characteristics of the Bailey Diagram,
we perform an analytic fit connecting the key parameters using polynomial regression, a machine learning algorithm based on supervised learning.
First, we choose the period $P$ and amplitude $A_V$ of RRLs as the independent and dependent variables, respectively, and split them into training and testing sets.
Then we pre-process the data and apply the polynomial-regression algorithm to the data set and build the model. 
Following that, we use the training data to fit the model and find the regression coefficients. We need to choose the $R^{2}$ (which lies in the range between $0$ and $1$, and measures the degree of fit for a model to the data; the closer to $1$, the better the model can approximate the data) to evaluate the performance of the model using test data, and adjust the model based on the evaluation results. 
Finally, we use the trained model to make predictions based on the data, and obtain the predicted values of the dependent variable. 
Through the above steps,
we determine the correlation between period and amplitude for RRab and RRc stars.
The analytic relations are as follows:

\begin{eqnarray} \label{equ}
	%{\rm RRab:\:} A_V = -1.45 - 12.29 \cdot \log{P} - 13.62 \cdot \log{P^2},
     {\rm RRab:\:} A_V = -1.45 (\pm 0.082) - 12.29 (\pm 0.713) \cdot \log{P} \nonumber \\ - 13.62 (\pm 1.501) \cdot \log {P^2},
\end{eqnarray}

\begin{eqnarray} \label{equa}
	%{\rm RRc:\:} A_V = -12 - 43.04 \cdot \log{P} - 35.46 \cdot \log{P^2}.
    {\rm RRc:\:} A_V = -12 (\pm 4.604) - 43.04 (\pm 19.918) \cdot \log{P} \nonumber \\ - 35.46 (\pm 21.046) \cdot \log{P^2}.
\end{eqnarray}
The uncertainties of the coefficients in the polynomial models are properly estimated from the covariance matrix.
We note that the luminosity amplitude is taken from two different sources. 
For 1857 stars we use the {\it G}-band time-series photometry from {\it Gaia} EDR3 \citep{GaiaCollaboration2021}. 
The {\it G}-band amplitude is then transformed into a {\it V}-band amplitude using equation 2 of \citet{2019Gaia}. 
For the remaining 1796 stars with poor {\it Gaia} phase coverage, the {\it V}-band amplitude is collected from literature sources \citep{vivas2004quest,watkins2009substructure,sesar2009light,mateu2012quest,drake2013probing,drake2014catalina}.

\citet{fabrizio2019use} produced a histogram of $A_V$, $\log{P}$ for stars in their RRab sample. They traced the local maxima and minima between the two main over-densities as the Oosterhoff-intermediate loci, and obtained a relation \citep{fabrizio2021use}\footnote{To separate RRab variables into Oo\,I and Oo\,II classes they adopted the following relation: $A_V=-1.39(\pm 0.09) - 13.76(\pm 0.97)\cdot\log P-15.10(\pm 2.64)\cdot\log P^2$ $[\sigma=0.19{\rm\,mag}].$} 
for dividing Oosterhoff groups. They did not provide an explicit relation between $A_V$ and $\log{P}$ for RRc stars. 
Fig.\,\ref{fig.Bailey} compares our relation with that from \citet{fabrizio2021use}. The grey solid line represents the division for RRab stars from \citet{fabrizio2021use}, which we shift by 0.06 units to the left to better fit the dichotomy of our sample. We note that this small shift is within the observational error, and it is possibly caused by the different metallicity and period distributions of the adopted samples.  
%(\textbf{due to the variations in sample size across different periods and metallicity ranges, we shifted the dividing line of Fabrizio to the left by 0.06 units (within the error range) to better exhibit dichotomy feature and our data}).
The red dotted line is also a RRab division line, but obtained from polynomial regression. The consistency of the results obtained by the two different methods suggest that polynomial regression is a feasible method to separate the Oosterhoff groups. We employ the same method to separate RRc stars into two groups.
The RRc variables are more sensitive to metallicity compared with RRab variables \citep{fabrizio2021use}, which leads RRc stars to exhibit a smooth transition over the period range, and thus they are difficult to separate accurately.
%even rather than a visible ``valley" on the Bailey diagram.

\section{Results}
\label{sec:result}

We identify 3653 RRLs (2661 RRab, 992 RRc) in the Galactic halo. The numbers of stars we assign to Oo\,I and Oo\,II groups for RRab\,/\,RRc types are 1899\,/\,524 and 762\,/\,468, respectively, as derived by polynomial regression (Equations $\ref{equ}-\ref{equa}$).
In this section, we use the period distribution, metallicity distribution, velocity distribution, anisotropy parameter $\beta$, orbital parameters, and action space of our sample to explore differences between the Oosterhoff groups and possible causes of the Oosterhoff dichotomy.

In order to quantify the differences between the Oosterhoff groups, we employ a 
two-sample Kolmogorov-Smirnov (KS) test to estimate the significance of the differences between Oo\,I and Oo\,II stars.
The KS test returns the probability $p_\mathrm{ks}$ under the null hypothesis that they have identical parent populations; the null hypothesis is rejected if $p_\mathrm{ks}$ $<$ 0.05.
We adopt the bootstrapping technique to create multiple simulated samples. After re-sampling 80\% of our Oosterhoff samples, and perform the KS test and iterating 1000 times, the resulting $p_\mathrm{ks}$  value was obtained from the mean of the resulting distribution. 

%The $p_\mathrm{ks}$ value will be the 
%criterion for determining whether two Oosterhoff group from the identical %distribution\ref{sec:intro}.

\subsection{Period Distribution}\label{subsec:period}

We first consider the mean periods for the two types of RRLs and their Oosterhoff groups. The mean period for RRab\,/\,RRc stars is $0.58\,/\,0.33$ days, respectively. For Oo\,I\,/\,Oo\,II RRab stars, the mean period is $0.55\,/\,0.64$ days, respectively. For the Oo\,I\,/\,Oo\,II RRc stars, the mean period is $0.31\,/\,0.36$ days, respectively. These results are consistent with the literature referred to in the Introduction.

\subsection{Metallicity Distribution}\label{subsec:feh}

Fig.\,\ref{fig.feh} compares the metallicity distribution of the Oo\,I and Oo\,II groups for our RRab and RRc samples.
For RRab stars, we find the mean metallicity of Oo\,I type stars is $\langle {\rm [Fe/H]} \rangle = -1.59$; the Oo\,II type stars have a more metal-poor mean value, $\langle {\rm [Fe/H]} \rangle = -1.87$. 
The means of the metallicity distribution for RRc stars (Oo\,I types: $\langle {\rm [Fe/H]} \rangle = -1.65$; Oo\,II types: $\langle {\rm [Fe/H]} \rangle = -1.96$) is more metal poor than for the RRab stars. 
We note that the abrupt drop around [Fe/H] = $-1$ for both RRab and RRc stars are not caused by our metallicity cut ([Fe/H]\,$< -1.0$). This drop is also noted by previous studies \cite[e.g.,][]{fabrizio2019use}.
The metallicity differences between RRab and RRc stars mainly arises because the morphology of RRc stars on the HB is bluer and hotter than that of RRab stars, which also supports previous ideas that the distribution of stars along the HB mainly depends on metal content \citep{renzini1983current,Torelli2019HorizontalBM}.

Early studies proposed that the Oosterhoff dichotomy in globular clusters is mainly a consequence of the deficit of metal-intermediate MW globular clusters hosting RRLs \citep{renzini1983current,castellani1983old}. They found that 
blue-HB clusters with [Fe/H] $\approx -1.8$ fill the gap between these types, and thus attributed the metallicity gap to the HB behaviour.
Since the discovery of the Oosterhoff dichotomy, many attempts have been made to find relationships between period, amplitude and metal abundance \citep{kunder2009oosterhoff,smith2011rr,fabrizio2019use,Bono2020OnTM} in order to explain this phenomenon from the perspective of stellar evolution \citep{sandage1981evidence,sandage1981oosterhoff,lee1990horizontal,clement1999rr,smith2011rr}. 
More recently, 
\citet{fabrizio2019use} collected a large data set of field RRLs, and found that both period and amplitude display a linear anti-correlation with metallicity -- an increase in the metal content causes a steady decrease in pulsation period and visual amplitude \cite[see figure 9 of][]{fabrizio2021use}. Thus, we cannot neglect the impact of stellar evolution on the Oosterhoff dichotomy.

\subsection{Velocity Distribution}\label{subsec:velo}

We now consider the space velocities of the Oosterhoff groups.
Fig.\,\ref{fig.vrvphi} shows the radial and rotational velocity distributions. Both groups centre around $V_\phi$ $\simeq$ 0. 

We can quantify the difference of the cylindrical velocity distributions between the Oosterhoff groups. The KS test comparing each of the Oo\,I and Oo\,II velocity components ($V_r$, $V_\theta$, $V_\phi$) for RRab stars
yields $p_\mathrm{ks}$ of (0.045, 0.002, 0.001), respectively.
The same test for the Cartesian velocity components $(U,V,W)$ yields $p_\mathrm{ks}$ of (0.536, 0.001, 0.123), respectively. 
The velocity components of RRc stars are also tested in the same way. The $p_\mathrm{ks}$ of the cylindrical velocity components and the Cartesian velocity components are (0.126, 0.001, 0.013) and (0.184, 0.025, 0.123), respectively. 
Statistically, the $V_\theta$ and $V_\phi$ components for the both RRab and RRc stars all reject the null hypothesis; these components differ between Oo\,I types and Oo\,II types.

\subsection{Anisotropy Parameter}\label{subsec:aniso}

To further investigate the dynamical properties of the Oosterhoff groups,
we describe the orbits using the anisotropy parameter $\beta$:
\begin{eqnarray}
	\beta(r) = 1 - \frac{{\sigma_{\theta}(r)}^2 + {\sigma_{\phi}(r)}^2}{2 \sigma_{r}(r)^2},
\end{eqnarray}
which represents the shape of the velocity ellipsoid as predominantly radial, tangential, or isotropic. The velocity dispersions ($\sigma_{\theta}$, $\sigma_{\phi}$, $\sigma_{r}$) are in the Galactocentric spherical polar coordinate system. As $\beta$ represents the ratio of tangential to radial motion, and measures the nature of orbits in the halo, it places restrictions on the merger history of the MW \citep{Rashkov2013,bird2015fad,hattori2017reliability,Belokurov2018CoformationOT,Loebman2017BetaDI,Lancaster2018TheHA}.

In order to check for differences between the Oosterhoff groups using the $\beta$ parameter, we divide our sample into seven bins according to Galactocentric distance, from 3 to 78 kpc in 5 kpc bins, with the first ($3 \leq r < 10\,$kpc) and the last two ($30 \leq r < 40\,$kpc, $r\geq 40\,$kpc) bins being exceptions. 
We obtain the mean velocity dispersion and errors for the individual velocity ellipsoid components using the bootstrap technique. We re-sample 80\% of the sample for each bin, calculate the mean velocity and dispersion, and repeat 1000 times. The velocity dispersion $\sigma$ and its error for each bin are the mean and standard deviation of the distribution yielded by re-sampling. The intrinsic velocity dispersion for each bin is a combination of two estimates: one is from the estimated velocity dispersion from re-sampling; the other is based on the the mean velocity uncertainty.

\begin{figure}
	\centering
	\subfigure
	{
		\begin{minipage}[b]{0.5\textwidth}
			\includegraphics[scale=0.49]{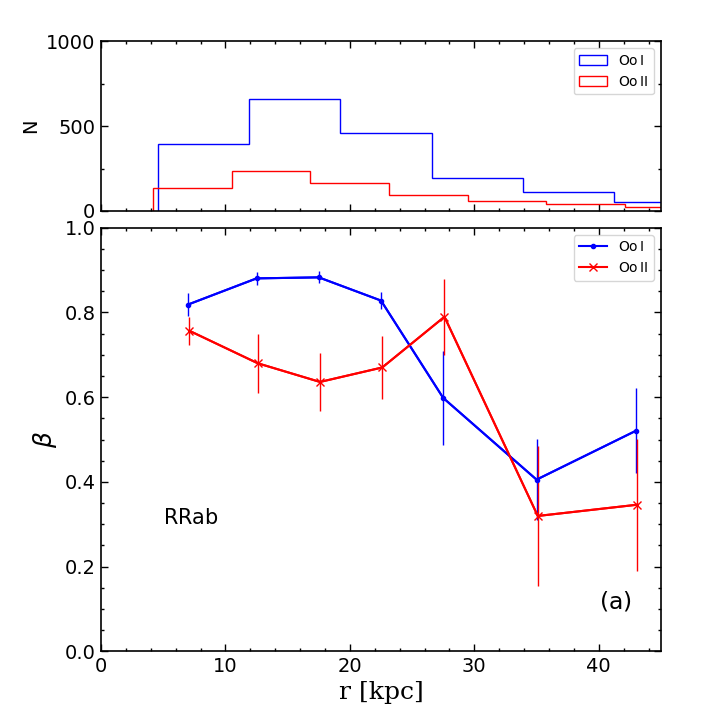}
			%\setlength{\belowcaptionskip}{-10cm}
			%\vspace{-0.5cm}
		\end{minipage}
	}
	\vskip -0.1cm
	\subfigure
	{
		\begin{minipage}[b]{0.5\textwidth}
			\includegraphics[scale=0.49]{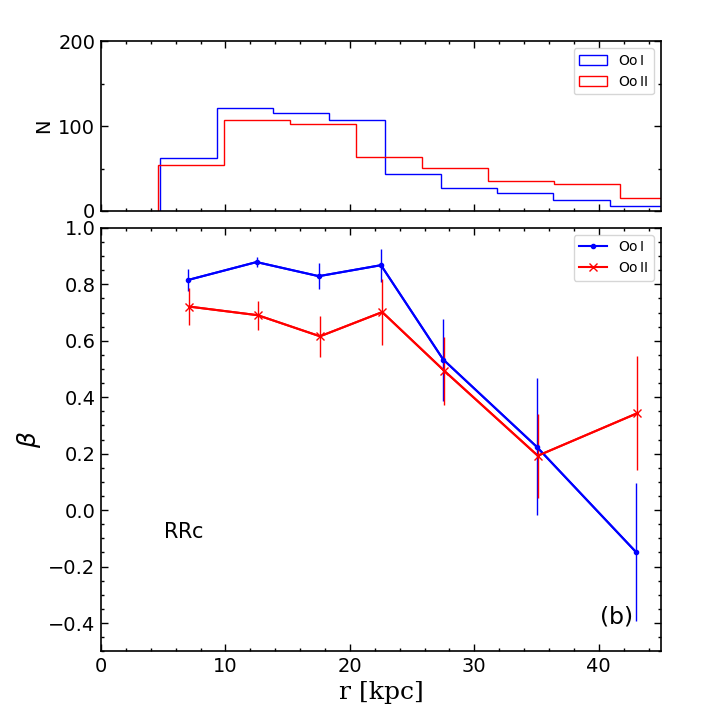}
		\end{minipage}
	}
	\caption{Velocity anisotropy $\beta$ trends and corresponding uncertainties as a function of Galactocentric distance $r$. Panel (a): The blue and red lines represent the trends in $\beta$ for Oo\,I and Oo\,II RRab stars, respectively. 
    The top histograms show the distribution of $r$ for the Oo\,I stars (blue) and Oo\,II stars (red).
    Panel (b): The same as in the upper panel, but for RRc stars.}
	\label{fig.beta}
\end{figure}

Fig.\,\ref{fig.beta} profiles the velocity anisotropy $\beta(r)$ as a function of Galactocentric radius for RRab and RRc stars. 
The Oo\,I RRab stars in panel (a) have high $\beta$ values, nearly $\sim0.9$, at $10$ kpc, remain almost the same to $20$ kpc, then drop past $25$ kpc. Within the same distance range where the IHR transitions to the OHR (25$-$30 kpc), the profile continues to decline. 
%and gradually increases when in outer halo ($r\geq 40\,$kpc). 
For Oo\,II RRab stars, $\beta$ declines in a relatively flat and continuous manner, and then increases to nearly $0.8$ at $25$ kpc. At the distance where the IHR and OHR transition occurs $\beta$ drops rapidly.
Panel (b) presents the trend of $\beta(r)$ for RRc stars. From inspection, the RRc stars do not exhibit a distinct change in $\beta$ between the Oosterhoff groups in the IHR compared with the RRab stars. One possible reason is that it is due to the mixture of the two populations, given the current measurement errors. This is particularly true for Oo II RRc stars, given its limited sample size. 
Both Oo\,I and Oo\,II RRc stars exhibit high and  nearly constant $\beta$ values within 10 kpc, and the velocity anisotropy of Oo\,I type stars is systematically higher than that of the Oo\,II type stars before about 25 kpc. Then $\beta$ value drops quickly at the IHR\,/\,OHR transition radius. 
%For both Oosterhoff groups $\beta$ drops quickly especially where the inner and outer halo meet. 
Regardless of whether a star is classified as RRab or RRc, this result shows that the Oo\,I-group stars possess higher $\beta$ values than that of the Oo\,II-group stars in the IHR.

A number of previous studies have analysed the halo's velocity anisotropy as a function of metallicity and distance, and shown that the metal-rich components exhibit strong radial anisotropy ($0.6 \leq \beta \leq 0.9$) for the IHR stars, and a milder radial anisotropy ($\beta \sim$ 0.5) for the OHR stars \citep{Belokurov2018CoformationOT,bird2019anisotropy,bird2021,Iorio2020ChemokinematicsOT,Liu2022ProbingTG}.
The strongly radial components are regarded as arising from the debris cloud of the ancient massive merger GSE substructure, which not only hosts a large fraction of Oo\,I RRLs \citep{Belokurov2017UnmixingTG}, but its contribution to the Galactic halo also decreases with increasing Galactoentric distance \citep{Iorio2020ChemokinematicsOT}.

Our results are in agreement with the above studies. We
find that, between 5-25 kpc, the velocity anisotropy of Oo\,I stars is relatively high ($0.75 < \beta < 0.90$), while that of Oo\,II stars is relatively low ($0.65 < \beta < 0.75$).
The fractions of the two groups between 5-25 kpc are 70\% and 30\% for Oo\,I and Oo\,II types for RRab stars, respectively, while for RRc stars the fractions are 60\% and 40\%, respectively.  Given the high radial motions and high fraction of Oo\,I stars in the IHR, we conclude that the Oo\,I type is mainly a relic associated with the GSE merger event.

\subsection {Action Space} \label{subsec:orbital}

\begin{figure*}  
	\centering
	\subfigure
	{
		\includegraphics[scale=0.7]{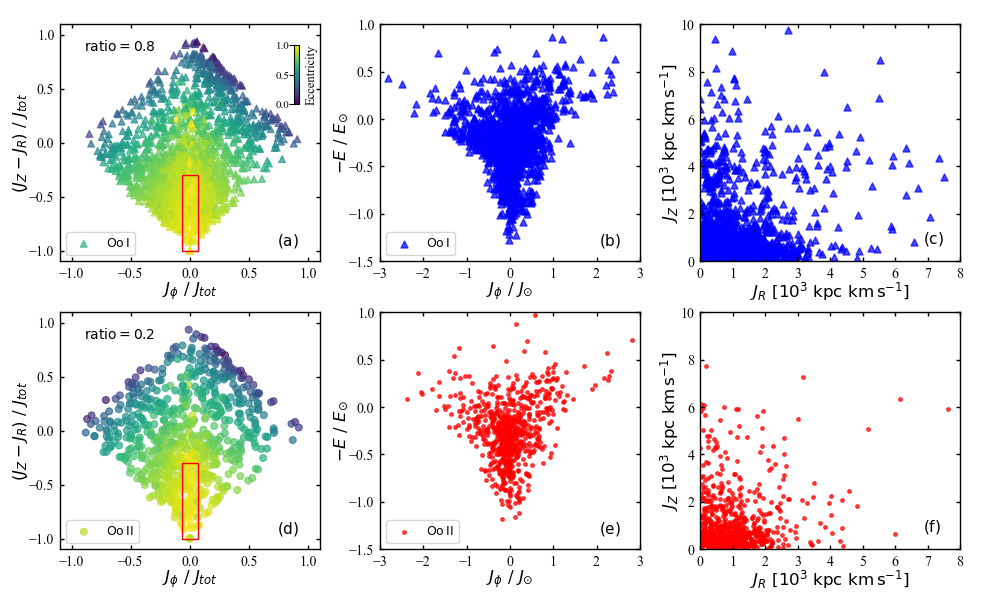}
		%\caption{fig1}
	}
	%	\quad
	%	\quad
	\subfigure
	{
		\includegraphics[scale=0.7]{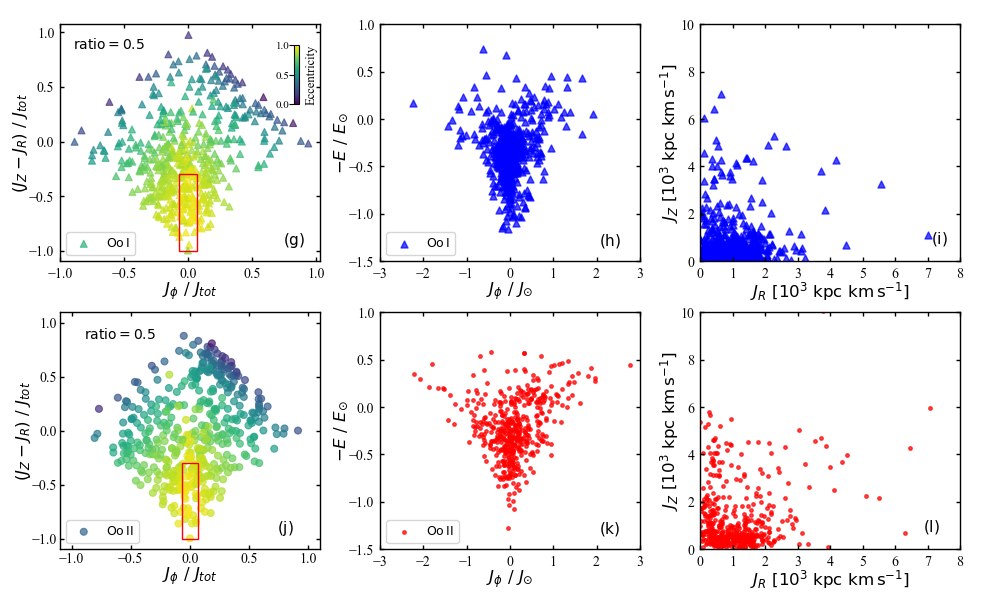}
	}
	
	\caption{Action and energy results for RRab stars (panels (a) - (f)) and RRc stars (panels (g) - (l)). Panels (a) - (c) show the results for Oo\,I stars. Panel (a) is the projected-action map for Oo\,I stars; each star is colour-coded by eccentricity as indicated by the colour bar. The red box encloses Oo\,I stars that lie in the GSE loci \citep{Myeong2019EvidenceFT}; ratios of Oosterhoff groups that lie in the GSE loci are also marked. Panel (b) shows the energy $(E)$ as a function of azimuthal action $(J_{\rm \phi})$, normalized by Solar values. Panel (c) shows the vertical action $(J_{\rm Z})$ as a function of radial action $(J_{\rm R})$. The second row in each set of panels are the same, but for Oo\,II stars.}
	\label{fig.action}	
\end{figure*}

The action-space coordinates for Galactic dynamics is considered an ideal plane to analyse halo stars, and identify substructures and debris from accretion events \citep{Binney1984SpectralSD}. 
Fig.\,\ref{fig.action} shows the distribution of Oosterhoff groups in action space; the upper panels (a) - (f) and lower panels (g) - (l) show the distribution for RRab and RRc stars, respectively.

Panel (a) and (d) show the action-space maps for Oo\,I and Oo\,II stars, respectively. The horizontal axis is the azimuthal action $J_\phi$; the vertical axis is the difference between the vertical and radial actions $J_Z-J_R$. Both axes are normalized by the total action $J_{tot} = J_{R}+J_Z+|J_\phi|$. Each star is colour-coded according to its eccentricity. 
\citet{Myeong2019EvidenceFT} identified that GSE-debris stars characteristically have $e \sim 0.9, |J_\phi\,/\,J_{tot}|$ and $(J_Z-J_R)\,/\,J_{tot} < -0.3$. We adopt the same membership criteria, and indicate GSE loci in the red box in panel (a) of Fig.\,\ref{fig.action}. We find that 80\% of Oo\,I RRab stars lie in the GSE loci. 
Panel (b) shows the distribution of energy vs. azimuthal action for Oo\,I stars, normalized by the Solar values ($J_{\phi, \odot}=2009.92\,\rm kpc\,\rm km\,s^{-1}$, $J_{Z, \odot}=0.35\,\rm \rm kpc\,\rm km\,s^{-1}$ 
and $E_\odot=-64943.61\,\rm km^2\,s^{-2}$; \citealt{Sestito2019TracingTF}). 
There is a compact distribution around $J_{ \phi} \simeq 0\,\rm kpc\,\rm km\,s^{-1}$, which
indicates an increased number of stars moving on radial orbits.
Retrograde orbits have negative $J_\phi$, while prograde orbits have positive $J_\phi$. Objects with higher-energy retrograde stars were merged in the past \citep{Myeong2018TheMW}.
Panel (c) presents the radial action vs. vertical action for Oo\,I stars. \citet{Myeong2018TheMW} analysed the local stellar halo in action space, and found that the metal-rich and high-eccentricity population is more extended toward higher radial actions, while the sample with radial orbits at low $J_{\rm R}$ is related to merger events.

\begin{figure*}
	\centering
	\begin{minipage}{0.49\linewidth}
		\centering
		\includegraphics[scale=0.54]{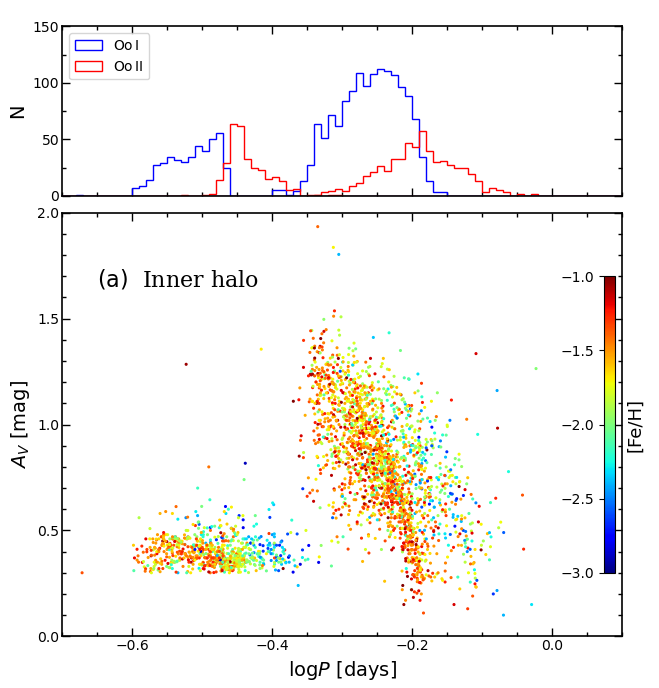}
		\label{lmc}
	\end{minipage}
	\begin{minipage}{0.49\linewidth}
		\centering
		\includegraphics[scale=0.54]{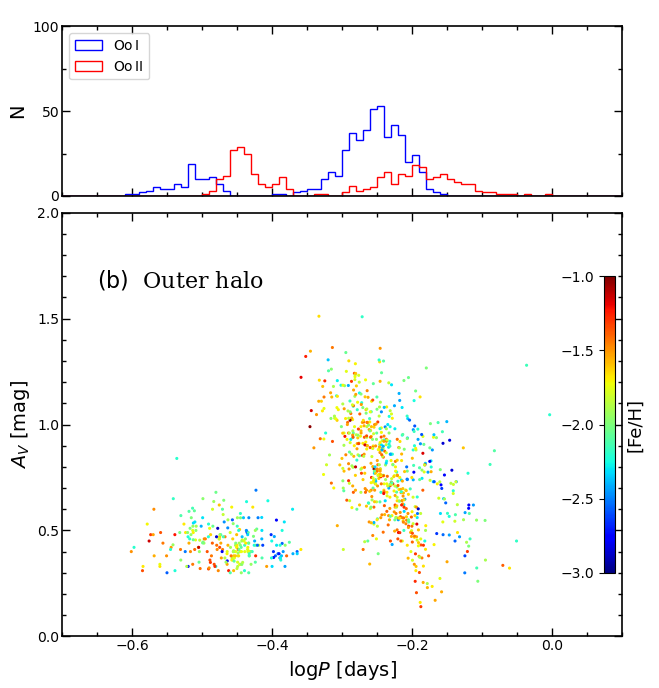}
		\label{smc}
	\end{minipage}
	%\qquad
	%让图片换行，
	
	\begin{minipage}{0.49\linewidth}
		\centering
		\includegraphics[scale=0.54]{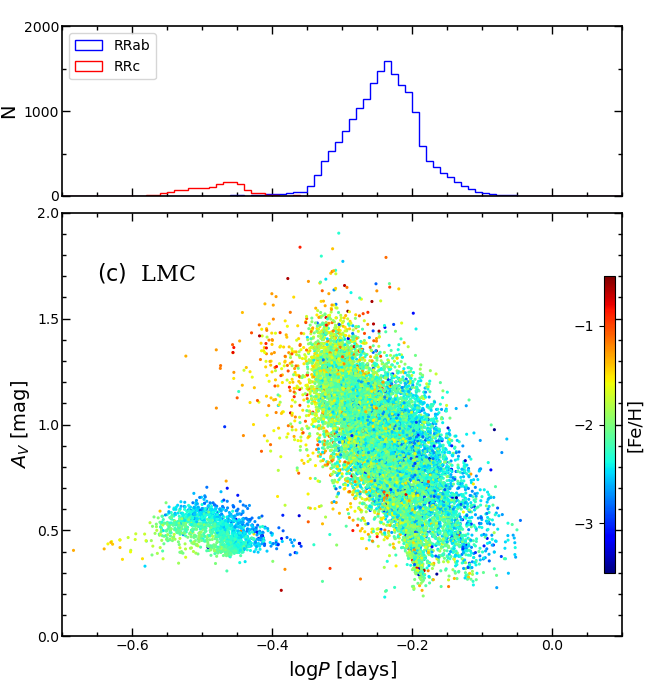}
		\label{inner halo}%文中引用该图片代号
	\end{minipage}
	\begin{minipage}{0.49\linewidth}
		\centering
		\includegraphics[scale=0.54]{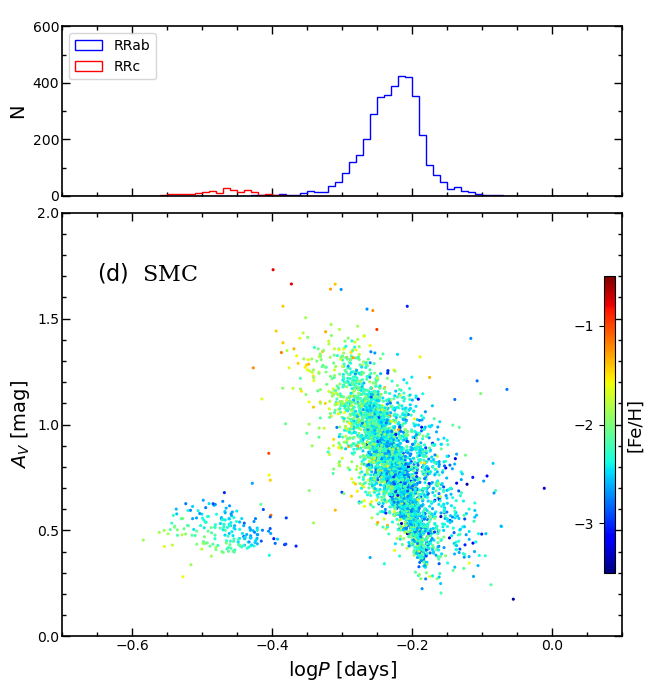}
		\label{outer halo}%文中引用该图片代号
	\end{minipage}
	\caption{Panels (a) and (b) are Bailey Diagrams for the MW IHR ($r<25$ kpc) and OHR ($r>25$), respectively. 
    The top histograms show the distribution of logP for the Oo\,I stars (blue) and Oo\,II stars (red).
    Panels (c) and (d) are Bailey Diagrams for the LMC and SMC, respectively. 
    The top histograms show the distribution of logP for the RRab stars (blue) and RRc stars (red).
    All sample stars are colour-coded by metallicity as indicated by the colour bar.}
	\label{fig.mc}
\end{figure*}

Panels (d) - (f) in Fig.\,\ref{fig.action} show the same plots, but for RRab stars belonging to the Oo\,II group.
From panel (d) we calculate that 20\% of Oo\,II RRab stars lie in the GSE loci. From the panels (b) and (e) of the Oo\,I and Oo\,II groups, respecively, we find that the relative fraction of Oo\,I stars occupying the high-energy retrograde region is much higher than for stars in the Oo\,II group. From panel (f) for Oo\,II stars, our result suggests that Oo\,II stars exhibit a lower spread toward highly radial actions, and are nearly evenly distributed in the three actions, while the Oo\,I stars in panel (c) have a tendency to have high $J_{R}$. The distribution of Oo\,I stars for $J_{R}$ is broader than that for Oo\,II-group stars, indicating that stars in the Oo\,I-group stars are a flattened population, while the stars in the Oo\,II group are from a rounder population.

We perform the same analysis as above for the RRc stars, and show the results in the lower set of panels (g) - (l) in Fig.\,\ref{fig.action}. The Oo\,I and Oo\,II dichotomies are not as evident for RRc stars as for RRab stars; this may be due to their smaller numbers. 
From the action-space analysis, we know that the Oo\,I stars are relatively metal rich, have highly radial actions, and are mostly distributed around $J_{\phi} \simeq 0\,\rm kpc\,\rm km\,s^{-1}$.  The Oo\,II stars are relatively more metal poor, and featured no obvious radial action compared with Oo\,I; they are more evenly distributed in the three actions.

Taking the analyses carried out in this section as a whole, we find that there exist significant differences between the Oosterhoff groups in their metallicity distributions, anisotropy parameters, and in action space. 
The Oo\,I group is associated with a relatively a more metal-rich component, with a strong trend towards radial motions and high eccentricity, zero angular momentum, and high-energy retrograde orbits. 
The Oo\,II group is associated with a relatively more metal-poor component, which has mild radial motions, and lower-eccentricity prograde orbits.
The features of the Oo\,I stars are in accord with those of GSE, which is thought to be the last major merger experienced by the MW about 10 Gyr ago. Many studies have investigated the GSE substructure. and found it has stars with high-energy, eccentric radial orbits, and exhibits a high anisotropy, reaching up to $\beta\ = 0.9$ 
\citep{Belokurov2018CoformationOT,Helmi2018TheMT,feuillet2020skymapper,Liu2022ProbingTG}.
We conclude that the Oosterhoff groups have different star-formation histories, with Oo\,I stars originating from the GSE substructure, while the Oo\,II stars formed $in-situ$ at an early epoch.

\begin{figure*} 
	\centering
	\subfigure{
		\includegraphics[scale=0.7]{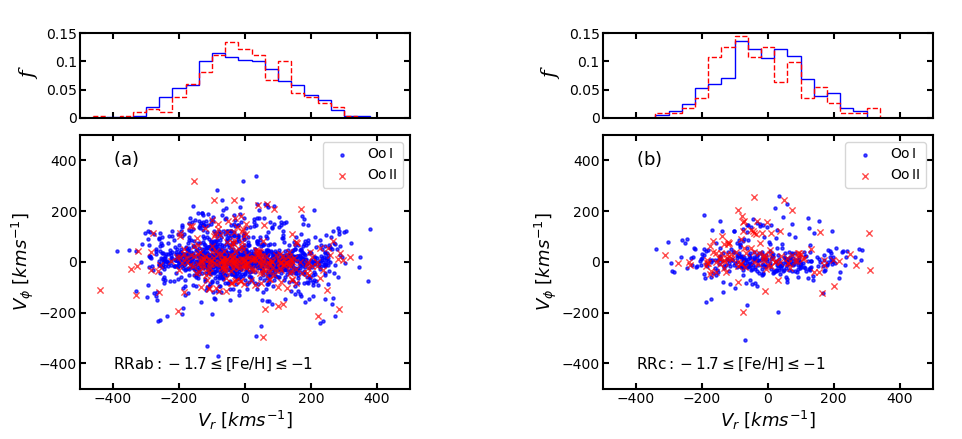}
		%\caption{fig1}
	}
	%	\quad
	%	\quad
	\subfigure{
		\includegraphics[scale=0.7]{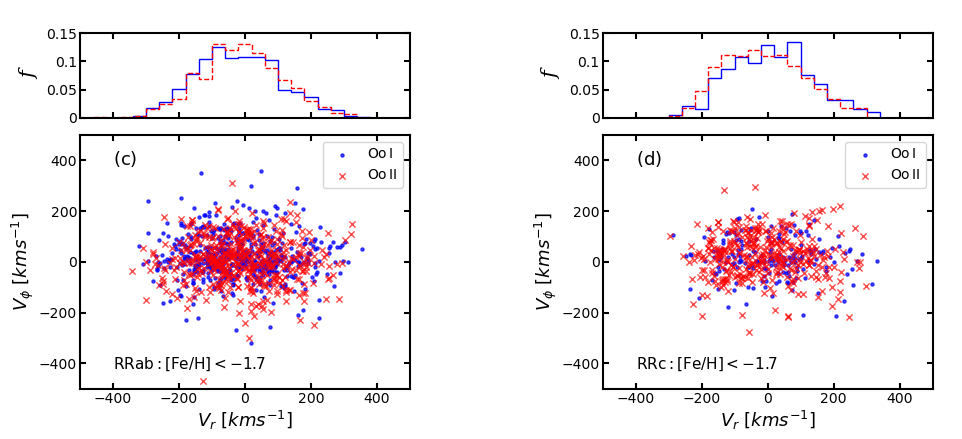}
	}
	
	\caption{Velocity ellipsoid distribution in Galactocentric spherical coordinates for Oo\,I stars (blue points) and Oo\,II stars (red crosses). Panels (a) and (b) show the results for stars with [Fe/H] $\geq -1.7$.  Panels (c) and (d) are for stars with [Fe/H] $< -1.7$.  RRab stars are shown in panels (a) and (c),  while the RRc stars are shown in panels (b) and (d).  The top histograms show the distribution of $V_r$ for the Oo\,I stars (blue) and Oo\,II stars (red).}
	\label{fig.susage shape}	
\end{figure*}

\begin{figure*}
	\centering
	\subfigure
	{
		\begin{minipage}[b]{.5\linewidth}
			\centering
			\includegraphics[scale=0.5]{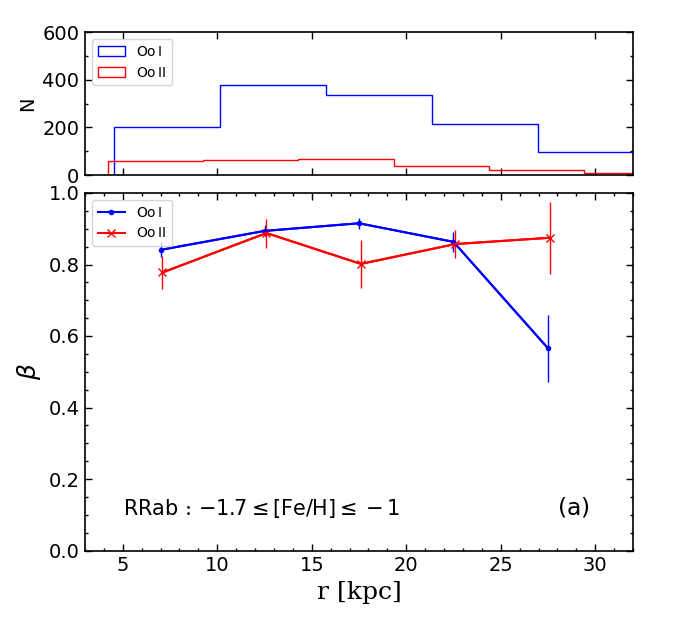}
		\end{minipage}
		\begin{minipage}[b]{.5\linewidth}
			\centering
			\includegraphics[scale=0.5]{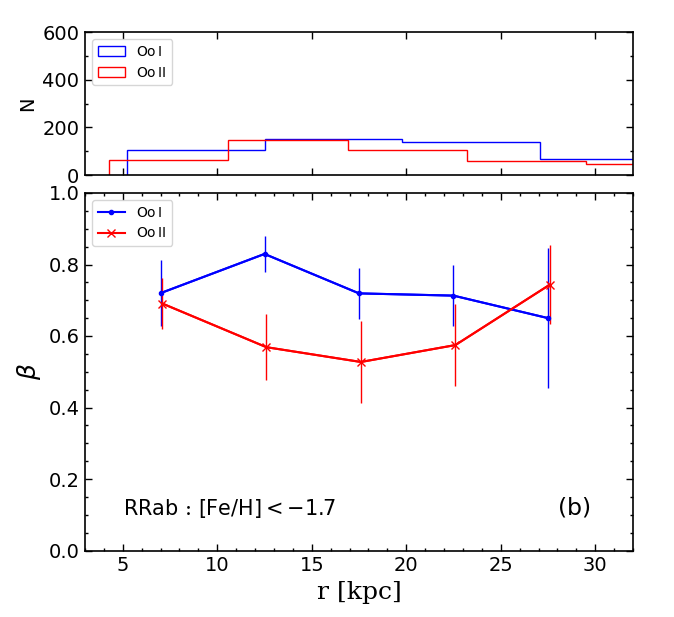}
		\end{minipage}
	}
	\vskip 0.1cm
	\subfigure
	{
		\begin{minipage}[b]{.5\linewidth}
			\centering
			\includegraphics[scale=0.5]{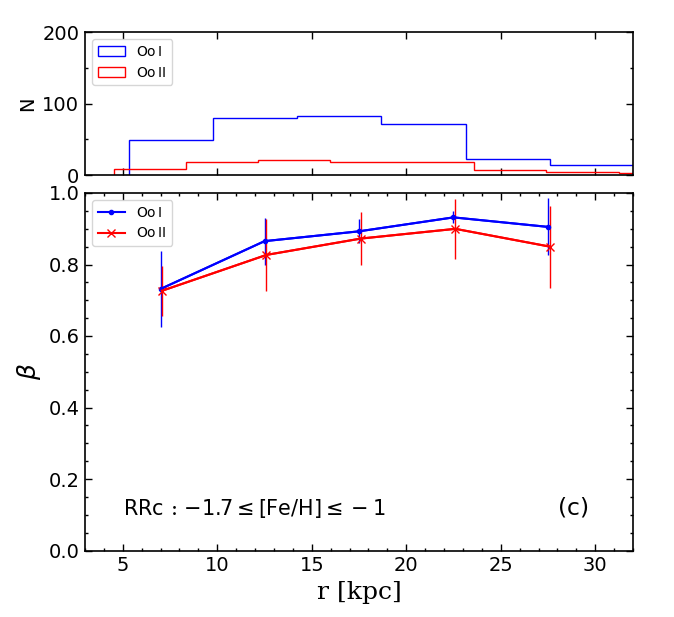}
		\end{minipage}
		\begin{minipage}[b]{.5\linewidth}
			\centering
			\includegraphics[scale=0.5]{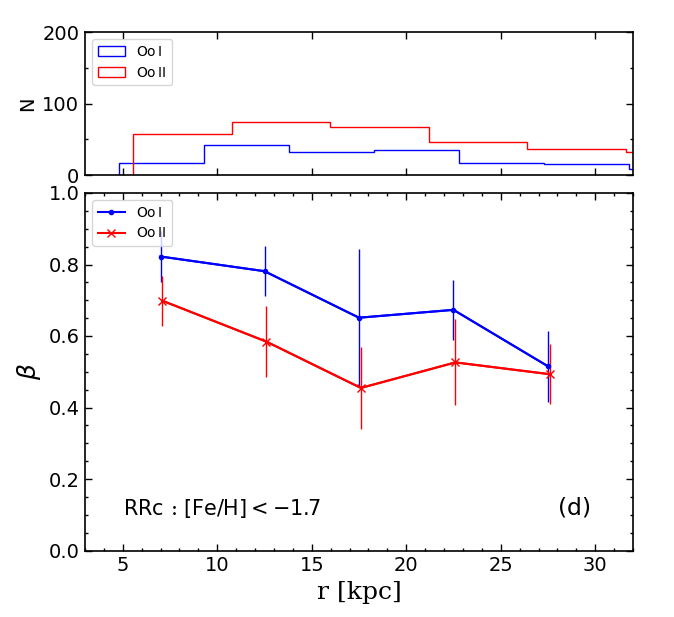}
		\end{minipage}
	}
	\caption{Velocity anisotropy parameter $\beta$, as a function of Galactocentric radius, for Oo\,I stars (blue) and Oo\,II stars (red) in different metallicity bins. Panels (a) and (b) show the trends for RRab stars. Panels (c) and (d) show the trends for RRc stars.  Panels (a) and (c) are for stars with [Fe/H] $\geq -1.7$, while panels (b) and (d) are for stars with [Fe/H] $< -1.7$. The top histograms show the distribution of $r$ for the Oo\,I stars (blue) and Oo\,II stars (red).}
 	\label{fig.anisotropy}
\end{figure*}

\section{Discussion}
 \label{sec:dis}
\subsection {The Inner-Halo / Outer-Halo Dichotomy} \label{Milky Way halo}

Does the Oosterhoff dichotomy exhibit variations based on location in the MW halo?
We take $r = 25$ kpc as the division to compare the Oosterhoff dichotomy in the IHR and OHR, shown in panels (a) and (b) of Fig.\,\ref{fig.mc}, respectively. 
It is clear that, in the IHR, Oo\,I-group stars with metal-rich components occupy the SP region, while the Oo\,II-group stars with metal-poor components are located in the LP region. In the OHR, the dichotomy is not as evident as in the IHR.

The IHR has a large contribution from the stellar debris associated with GSE \citep{Belokurov2018CoformationOT,myeong2018sausage,Helmi2018TheMT,Lancaster2018TheHA}, which explains why the the dichotomy is more apparent than in the OHR.
\citet{Liu2022ProbingTG} analysed the kinematics and chemistry of the Galactic halo RRLs, and also found that the inner-halo population mainly comprises stars deposited from this ancient merged satellite, and the outer-halo population exhibits only a mildly radial anisotropy.
The influence of GSE results in a buildup in the number of stars in the metal-rich regions, and we can observe a valley in the transition region.
The outer halo does not change so significantly, because it is not affected by the GSE, so it exhibits a gradual transition.
In addition, we have noticed that the proportion of stars in the Oo\,I and Oo\,II groups varies between the inner and outer halos. In the IHR, the ratio of Oo\,I-group stars is significantly greater than that of Oo\,II-group stars, while in the OHR, the ratio of Oo\,I and Oo\,II stars are roughly equal. Hence, the Oosterhoff dichotomy is more apparent for the IHR than for the OHR.

\subsection {The Oosterhoff Dichotomy of Local Group Dwarf Galaxies} \label{lmc}
Does the Osterhoff dichotomy exist in MW satellites, such as Fornax, Sagittarius, and the Large and Small Magellanic Clouds (LMC and SMC)?
 \citet{2009Astrophysics} found that the RRLs in dwarf galaxies and their globular clusters have intermediate periods, between the two Oosterhoff groups, and fill the gap with $0.58$ $< P < 0.62$ days. 
\citet{smith2011rr} found that RRLs in dwarf galaxies around the MW appeared to be genuinely 
Oosterhoff-intermediate, rather than a mixture of Oo\,I- and Oo\,II-group stars.
\citet{Soszynski2016TheOC} explored five LMC clusters, and validated that the Oosterhoff dichotomy observed in Galactic globular clusters is not present among the globular clusters in this galaxy.

We explore the Magellanic Clouds in more detail, as they provide an important piece of the puzzle to help us better understand galaxy interactions, as well as the dynamical and star-formation history of our Galaxy \citep{TepperGarca2019TheMS,Wang2022LessonsFT}.
\citet{Soszynski2016TheOC} collected over 45,000 RRLs, of which 39,082 were detected towards the LMC and 6369 towards the SMC, during the fourth phase of the Optical Gravitational Lensing Experiment (OGLE-IV). From a cross-match with \citet{li2022photometric}, who presented more-precise metallicity and distance estimates for over 135,000 RRLs by calibrating the P-$\phi_{31}$-[Fe/H] and {\it G}-band absolute magnitude-metallicity relations, we obtain a final sample including 19,078 RRLs in the LMC and 4022 RRLs in the SMC. 

Panels (c) and (d) in Fig.\,\ref{fig.mc} show the Bailey Diagrams for the LMC and SMC, respectively. Neither exhibit as obvious an Oosterhoff dichotomy as seen in the MW halo, which shows a clear metallicity gradient when moving from long periods to short periods at a fixed amplitude; they are more mixed together.
In the metal-abundance\,/\,period transition region, the LMC and SMC appear continuous, without the valley region that appears in the MW halo.
The LMC and SMC are located far from the Galaxy (48.84 kpc and 61.38 kpc away from the Galactic Centre, respectively; \citealt{li2022photometric}), and were likely unaffected by the GSE merger event in the MW.  Rather, they are influenced by the environment in each galaxy. By comparing the Bailey Diagrams of the LMC and SMC, we clearly see that LMC shows relatively continuous variations in both its period and metallicity, without any obvious dichotomy, and appears to be a gradual transition.
%By comparing the Bailey Diagrams of the LMC and SMC, we can clearly see that the variation of LMC is relatively continuous, without any obvious dichotomy, and appears to be a natural transition.
There are more metal-poor stars in the SMC, and we observe a distinct difference in the nature of the dichotomy with respect to that found in the LMC.

This confirms our conclusion that the Oosterhoff dichotomy is a complex phenomenon that varies across different galaxies, and even within different locations in the same galaxy, due to the fact that each location in different galaxies has undergone a unique star-formation history.

\subsection {Different Metallicity Bins}

\cite{Myeong2018TheMW} proposed that the kinematics of the metal-rich halo stars with [Fe/H] between $-1.9$ and $-1.1$ provides evidence for recent accretion events, and that the Oosterhoff dichotomy itself is correlated with metallicity \citep{smith2011rr}. To investigate the dependence of the nature of the observed Oosterhoff dichotomy on metallicty, and the cause of the metallicity differences between the groups, we divide our sample at [Fe/H] = $-1.7$, and consider the kinematic characteristics of the individual Oosterhoff groups. 

Fig.\,\ref{fig.susage shape} shows the velocity ellipsoids for the RRLs in two separate metallicity bins.  Panels (a) and (b) in this figure apply to RRLs with [Fe/H] $\geq -1.7$, which exhibit a remarkable ``sausage-like'' structure.  Both Oosterhoff groups are centreed around $V_\phi$ $\simeq$ 0 and are extended in $V_r$.
\cite{Belokurov2018CoformationOT} initially pointed out the GSE substructure in the $V_\phi {\rm vs.} V_r$ space; \citet{Helmi2018TheMT} concluded that it is the result of a head-on collision with the MW that deposited stellar debris on highly radial orbits, giving rise to the characteristic shape of the velocity ellipsoid, which becomes weaker and rounder with decreasing metallicity.
As seen in panels (c) and (d) of this figure, for the RRLs with [Fe/H] $< -1.7$, the strong radial character of the distribution is not as evident as for the more metal-rich sample.

Fig.\,\ref{fig.anisotropy} shows the variation of the anisotropy parameter $\beta$ with Galactocentric radius $r$ from 3 to 30 kpc. The metal-rich RRL sample possesses a high $\beta$, especially around 15 kpc, and the anisotropy parameter for stars in the Oo\,I group is higher than for those in the Oo\,II group, as mentioned in \ref{subsec:aniso}. The top histograms in these panels show that the proportion of stars in the Oo\,I and Oo\,II groups. It is clear that stars in the Oo\,I group represent a much larger proportion than in the Oo\,II group in the metal-rich bins, while the proportion of Oo\,I and Oo\,II stars are nearly equal in the metal-poor bins. 
These differences between the kinematics and proportions of Oo\,I and Oo\,II stars for different metallicity bins confirm that the metal-rich component is dominated by stars from GSE, and that the differences in metal abundance are mainly due to the evolution of the Galaxy.

\subsection {Comparison with the Results of \citeauthor{fabrizio2019use}}

\citet{fabrizio2019use,fabrizio2021use} discussed the largest and most homogeneous spectroscopic dataset of 9015 RRLs (6150 RRab, 2865 RRc), and found a linear correlation between mean periods and metallicity, which strongly indicates that the long-standing problem of the Oosterhoff dichotomy among Galactic globular clusters is a consequence of the lack of metal-intermediate clusters hosting RRLs.

Our results, based on the large spectroscopic sample of halo RRls from \citet{Liu2020ProbingTG} and the photometric sample from \citet{2022Probing},
indicate that the Oosterhoff groups in the MW have different kinematical and dynamical properties, which largely correspond to their different origins. 
From newly measured metallicity and distance estimates for RRLs in the LMC and SMC, we find there is no clear Oosterhoff dichotomy in these dwarf galaxies; the Oosterhoff dichotomy is not the same in different galaxies.  
Meanwhile the studies of \citeauthor{fabrizio2019use} suggested that the mean period of RRL variables is a continuous function of metallicity.
All of these findings indicate that it not feasible to explain the Oosterhoff dichotomy based on stellar physical mechanisms alone.

We also want to clarify differences in our reported fractions compared with the work of \citeauthor{fabrizio2019use}. We calculate the relative proportions of RRLs in the Bailey Diagrams according to a division line obtained from polynomial regression. The proportions of RRab stars in the the SP sequence and the 
LP sequence are 70\% and 30\%, respectively. The RRc stars have  proportions in the LP and SP sequences that are 50\% and 50\%, respectively.
In contrast, \citet{fabrizio2021use} used contours to obtain the sequence proportions. The RRab stars in the SP and LP sequences are 80\% and 20\%, respectively, comparable to our results. Their RRc stars exhibit an opposite trend between the SP and LP sequences, 30\% and 70\%, respectively. 
We believe this results from the different definitions of the SP and LP sequences.  We consider the $A_V-\log{P}$ correlation to separate the two sequences, while \citet{fabrizio2021use} separated them on the basis of density in their data space.
%Besides the different classification of sequence and we only have a small sample size of RRc variables.

\section{Summary} \label{sec:sum}

In this work, we employ 2661 RRab and 992 RRc variables with available 7D (3D position, 3D velocity, and metallicity) information to investigate the Oosterhoff dichotomy from the perspective of metallicity, velocity, anisotropy, and action space.

For RRab stars, both the Oo\,I and Oo\,II groups have high $\beta$ (0.85) over Galactocentric radii $5 < r < 25$ kpc; the proportion of Oosterhoff groups within $r = $ 25 kpc is 70\% for the Oo\,I stars and 30\% for the Oo\,II stars, respectively. For RRc stars within $r = 25$ kpc, the proportion is 60\% and 40\% for types Oo\,I and Oo\,II, respectively. 

From consideration of the action space, we find that stars in the Oo\,I group
have highly radial and eccentric orbits, and are mostly distributed around the null azimuthal-momentum region. The stars in the Oo\,II groupe have only mildly radial and eccentric orbits.
These results suggest that the Oosterhoff groups in the MW halo originated in different ways.  The Oo\,I group is associated with a relatively metal-rich component with highly radial, eccentric, and retrograde orbits and $\beta$ reaching values as high as 0.9 over $5 < r < 25$ kpc. Stars in the Oo\,I group likely originate from the GSE substructure, while stars in the Oo\,II group may have formed $in-situ$ during early epochs, and were later accreted.

The relative contributions of Oo\,I and Oo\,II stars in the Galactic halo varies with Galactocentric distance, and  the dichotomy in the IHR is more apparent than for the OHR. We have also inspected the nature of the Oosterhoff dichotomy in the LMC and SMC, and find differences in the phenomenon compared with the halo of the MW. 

We conclude that the Oosterhoff dichotomy is the result
of both stellar evolution and galactic evolution, and therefore varies with position and environment.

\section*{Acknowledgements}

Z.S. and L.G.C. acknowledge the Hubei Provincial Natural Science Foundation with grant No. 2023AFB577, the Key Laboratory Fund of Ministry of Education under grant No. QLPL2022P01 and the National Science Foundation of People's Republic of China (NSFC) with No. U1731108. H.Y. acknowledges the National Key R\&D Program of China No. 2019YFA0405503 and the NSFC with Nos.11903027 and 11833006. Z.H.W. acknowledges the National Key R\&D Program of China No. 2019YFA0405504 and the NSFC with Nos.11973001, 12090040 and 12090044. T.C.B. acknowledges partial support for this work from grant PHY 14-30152; Physics Frontier Center\,/\,JINA Center for the Evolution of the Elements (JINA-CEE), and OISE-1927130: The International Research Network for Nuclear Astrophysics (IReNA), awarded by the US National Science Foundation. T.H.J. acknowledges NSFC with No. 11873034 and the Hubei Provincial Outstanding Youth Fund No. 2019CFA087.

The Guoshoujing Telescope (the Large Sky Area Multi-Object Fiber Spectroscopic Telescope LAMOST) is a National Major Scientific Project built by the Chinese Academy of Sciences. Funding for the project has been provided by the National Development and Reform Commission. LAMOST is operated and managed by the National Astronomical Observatories, Chinese Academy of Sciences. This work has made use of data from the European Space Agency (ESA) mission {\it Gaia} (\url{https://www.cosmos.esa.int/gaia}), processed by the {\it Gaia} Data Processing and Analysis Consortium (DPAC, \url{https://www.cosmos.esa.int/web/gaia/dpac/consortium}). Funding for the DPAC has been provided by national institutions, in particular the institutions participating in the {\it Gaia} Multilateral Agreement. This work also has made use of data products from the SDSS, 2MASS, and WISE.

%\appendix
%\section{Priors of the parameters of the metallicity distribution model}

%The priors of the assumed parameters of the metallicity [Fe/H] distribution model of each radial bin are all uniform within the ranges listed in Table\,A1.

%%%%%%%%%%%%%%%%%%%%%%%%%%%%%%%%%%%%%%%%%%%%%%%%%%
\section*{Data Availability}

The data supporting this article will be shared upon reasonable request sent to the corresponding authors.


%\bibliographystyle{mnras}
%\bibliography{bibfile} % if your bibtex file is called example.bib

%\begin{thebibliography}{99}
%\end{thebibliography}

%%%%%%%%%%%%%%%%%%%%%%%%%%%%%%%%%%%%%%%%%%%%%%%%%%

%%%%%%%%%%%%%%%%% APPENDICES %%%%%%%%%%%%%%%%%%%%%

%%%%%%%%%%%%%%%%%%%%%%%%%%%%%%%%%%%%%%%%%%%%%%%%%%

% Don't change these lines
\bsp	% typesetting comment
\label{lastpage}
\end{document}